\documentclass[12pt]{article}
\usepackage{amsfonts,amssymb,amsmath,latexsym}

\newcommand{\be}{\begin{equation}} 
\newcommand{\ee}{\end{equation}} 
\newcommand{\bea}{\begin{eqnarray}} 
\newcommand{\eea}{\end{eqnarray}} 
 
\newcommand{\ba}{\begin{array}} 
\newcommand{\ea}{\end{array}} 
 
\newcommand{\bt}{\begin{tabular}} 
\newcommand{\et}{\end{tabular}}

\newcommand{\sump}{\mathop{{\sum}'}} 
\newcommand{\ti}{\tilde} 
 
\newcommand{\fr}{\frac} 
 
\newcommand{\cl}{\centerline} 
\newcommand{\bs}{\bigskip} 
\newcommand{\vs}{\vspace}

\newcommand{\bbib}{\begin{thebibliography}} 
\newcommand{\ebib}{\end{thebibliography}}

\newcommand{\mbb}{\mathbb} 
 
\newcommand{\bm}{\boldmath}
\newcommand{\mb}{\mbox}
\newcommand{\scrs}{\scriptsize}

\begin{document}

\titlepage

\centerline{\large {\bf TOPOLOGICAL PHASE TRANSITIONS}} 
 
\bigskip 
 
\centerline{\large {\bf IN TWO-DIMENSIONAL SYSTEMS WITH}}

\bigskip

\cl{\large {\bf INTERNAL SYMMETRIES}} 
 
\vspace{1cm} 
 
\cl{\bf S.A.Bulgadaev}
 
\vspace{0.5cm} 
 
\centerline{L.D.Landau Institute for Theoretical Physics} 
 
\centerline{Kosyghin Str.2, Moscow, 117334, RUSSIA} 
 
\vs{1cm} 
 
\centerline{A talk given at seminar}
\bs 
\cl{\large "Topological Defects in Non-Equilibrium Systems} 
\bs
\cl{\large and Condensed Matter"}
\bs
\cl{Dresden, 22 July, 1999, Germany}
 
\newpage 

\cl{\large {\bf CONTENT}}

\vs{2cm}

\hbox to 14cm{{I. Introduction}\dotfill 3}
\bs
\hbox to 6cm{II.  Multicomponent generalization and spaces with 
vector $\pi_1$..........................\dotfill  6}
\bs
\hbox to 14cm{{III. $\sigma$-models on $T_G$, duality and effective 
theories.......................................}\dotfill 10}
\bs
\hbox to 14cm{IV. Phase diagram, critical properties and symmetries...........
.......................\dotfill 13}
\bs
\hbox to 6cm{{V. d-dimensional conformal $\sigma$-model and 
topological excitations...................}\dotfill 19}
\bs
\hbox to 14cm{{VI. Conclusions}\dotfill 22}
\bs
\hbox to 14cm{VII. References\dotfill 22} 

\newpage

\cl{\large {\bf I. Introduction}}

\bs

The topological phase transition (TPT) 
or the Berezinskii-Kosterlitz-Thouless (BKT)   
phase transition (PT) takes place in two-dimensional 
systems with degenerated vacuum, whose  order parameter 
$\psi = e^{2\pi i \varphi} \in {\cal M}=S^1.$ 
Among them are $XY$-model, superconductors, bose-liquids and many other
systems. 

\bs

\underline {\bf The main unusual feathers of this TPT are:}

\bs

1) absence of the spontaneous magnetization in low-T phase (Stanley, Kaplan
1966);

2) correlation functions in low-temperature phase must fall off 
algebraically (Rice 1966). 

\bs

In particular, $XY$-model on a lattice is defined as follows
$$
{\cal Z}_{XY} = \sum \exp (-\beta {\cal H}), \quad 
$$
$$
{\cal H}= -  \fr{1}{2}\sum_{<i,j>} J (\psi_i \psi_j^* + c.c.)=
- J \sum_{<i,j>} \cos(\varphi_i - \varphi_j).
$$
Its continuous variant is a nonlinear $\sigma$-model (NSM) on $S^1$
$$
{\cal Z}_{NS}= \int D\varphi e^{-{\cal S}[\varphi]}, \quad
{\cal S}_{NS}= \fr{1}{2\alpha}\int |\partial \psi|^2 d^2 x = 
\fr{1}{2\alpha} \int (\partial \varphi)^2 d^2 x,
$$
$$
\alpha \simeq T/2J.
$$
This form can be also considered as a longwave approximation of the 
 Ginzburg-Landau (GL) theory 
$$
{\cal S}_{GL}=  \int d^2x \left[\fr{1}{2\alpha}|\partial \psi|^2 +
M^2(|\psi|^2-1)^2\right]
$$
with large $M.$

A circle $S^1$ has a discrete abelian
first homotopic group $\pi_1$  
$$\pi_1(S^1)=\mbb {Z}.$$
Due to this property the topologically stable excitations, vortices, 
are possible in these systems. 

One vortex solution has a form (at large distancies)
$$
\varphi({\bf x})= \fr{1}{\pi} \arctan \fr{y}{x} + \varphi_0, 
\quad {\bf x} =(x,y) \in \mbb {R}^2,
$$
where $\varphi_0$ is some constant. Its choice defines a form of
vortex. For $\varphi_0 = 0$ it is pure potential, and for $\varphi_0
= \pi/2$ it is pure rotational. 
%(fig.1)
%\begin{figure}[h]
%\begin{tabular}{|c|c|}
%\hskip4cm
%{\psfig{figure=charge.eps,width=5cm}
%\vskip0.5cm
%\caption{Vortex 1.}} &
%\end{figure}
%\begin{figure}[b]
%\hskip4cm
%{\psfig{figure=vortex.eps,width=5cm}
%\vskip0.5cm
%\caption{Vortex 2.}}
%\end{tabular}
%\end{figure}
An account of vortices means that theory must be considered on the
covering space $\mbb {R}$ of the circle $S^1 = \mbb {R}/\mbb {Z}.$

The energy of one "vortex" with topological charge $e\in \mbb {Z}$ 
is logarithmically divergent 
$$ 
E= \fr{e^2}{2\alpha} 2\pi \ln \fr{R}{a}, 
$$ 
where $R$ is a space radius, $a$ is some short-wave cut-off parameter
(a radius of the vortex core or a lattice constant).

The energy of $N$-"vortex" solution, $E_N,$ with the full topological 
charge $e = \sum _{i=1}^N e_i = 0$ is finite and equals 
$$ 
E_N= \fr{2\pi}{2\alpha}\sum_{i\ne k}^N e_i e_k 
\ln \fr{|{\bf x}_i-{\bf x}_k|}{a} + C(a) \sum_i^N e_i^2, 
$$ 
here $C(a)$ is some nonuniversal constant, determining "self-energy" 
(or core energy) 
of vortices and depending on type of core regularization. 

Berezinsky (1970,1971), Popov (1971,1976) 
and Kosterlitz and Thouless  (1973,1974) have shown 
for the first time an important 
role of these topological 
exitations  for existence of TPT in two-dimensional systems with continuous
abelian symmetry. 

These results have stimulated further investigations, given 
a discovery of monopoles (Polyakov 1974, t'Hooft 1974), 
instantons (Belavin et al. 1975), 
and others 
topological excitations in different branches of physics 
(Toulouse and Kleman 1976, Volovik and Mineev 1977). 

The partition function ${\cal Z}_{XY}$
can be 
approximated by product of two partition functions
(Kosterlitz 1974, Jose et al.1977)
$$ 
{\cal Z}_{XY}\simeq {\cal Z}_{sw} {\cal Z}_{CG},
$$ 
where ${\cal Z}_{CG}$  is the grand partition function
of dilute Coulomb gase (CG) of topological exitations with minimal 
charges $e=\pm1$ and 
${\cal Z}_{sw}$ is the partition function of
{\emph{free}} "spin-waves".
${\cal Z}_{CG}$  
can be represented, in its turn, in the form 
of effective  field theory with noncompact field $\phi$ and sine-Gordon 
action ${\cal S}_{SG}$ 
(Jose et al. 1977, Wiegmann 1978)
$$
{\cal Z}_{CG}= {\cal Z}_{SG} = \int D\phi e^{-{\cal S}_{SG}[\phi]} ,
$$ 
$$
{\cal S}_{SG}= \int \left[ \fr{1}{2\alpha}(\partial \phi)^2
- 2 \mu^2 \cos \phi\right] d^2 x
$$
The TPT
takes place in the system of  vortices and can be described by 
the effective SG theory.
This system has two different phases:

1) \underline {\bf high-T phase}: plasma-like, massive, with finite 
correlation length with essential singularity at $T_c$
$$
\xi \sim a\exp (A\tau^{-\nu}), \quad \nu=\fr{1}{2},
$$
$$  
\tau = \fr{T-T_c}{T_c} \to 0, \quad T_c \approx \pi J;
$$

2) \underline {\bf low-T phase}: dielectric, massless, with infinite 
correlation length and algebraically falling correlations.

\bs

The TPT in low-dimensional systems can describe such important 
physical phenomena  as compactification -
decompactification (Green et al., 1988), 
localization - delocalization (Langer et al. 1987, Schmid 1983, Bulgadaev 1984) 
coherence - decoherence (Zaikin and Schon 1990) and 
others. For this reason
it is very interesting to find the next

\bs

\underline {\bf Possible generalizations of TPT on:}

\bs 

1) multicomponent 2D systems, having more complex vacuum manifolds ${\cal M}$ 
with nontrivial $\pi_1({\cal M})$;

2)D-dimensional systems, admitting the interacting topological excitations
with logarithmically divergent energies.

\newpage
%\bs

\cl{\large {\bf II. Multicomponent generalization and }}

\bs

\cl{\large {\bf spaces with vector $\pi_1$}}

\bs

{\bf 1. Usual tori.}

\bs

The simplest generalization of circle 
$$S^1 = e^{2\pi i\varphi}, \quad 0\le \varphi \le 1$$ 
is a  torus $T^n,$ 
which equals to the direct product  of $n$ circles 
$$ 
T^n = {\mbb R}^n/{\mbb Z}^n = \bigotimes^n_{i=1} S^1_i,  
$$ 
where ${\mbb R}^n$ is $n$-dimensional 
euclidean space,
${\mbb Z}^n$ is $n$-dimensional simple cubic integer-valued 
lattice 
$$ 
{\mbb Z}^n = \bigoplus_{i=1}^n {\mbb Z}_i.  
$$ 
A torus $T^n$ can appear as vacuum manifold of superposition of 
$n$ independent abovementioned  GL theories
$$
{\cal S}= \sum_i^n {\cal S}_{GLi}
$$
It is important to note, that
a symmetry group of this theory is a semidirect product of $T^n$ and a 
finite discrete group of 
permutations instead of usual continuous unitary or orthogonal groups. 

The homotopy group of torus $\pi_1(T^n)$ is usually written as 
$$ 
\pi_1(T^n) = \bigoplus^n_{i = 1}{\mbb Z}_i = {\mbb Z}^n, 
$$ 
where $i$-th component describes maps of the boundary $S^1$ into 
$i$-th circle of $T^n.$ These circles form so called "bouquet" of circles 
(a set of circles with one common point)
$$B^1_n = S^1_1\vee ... \vee S^1_n.$$ 
%\bs

\begin{picture}(300,60)
\put(240,45){\circle{30}}
\put(240,30){\circle*{3}}
\put(130,30){\vector(1,0){50}}
\put(226 ,20){\circle{30}}
\put(254,20){\circle{30}}
\put(80,30){\circle{60}}
\put(270,30){$B_n^1$}
\end{picture} 

%\bs
It is clear that maps into different components cannot 
be transformed into or annihilate each other. 
Consequently, one can introduce in $\pi_1(T^n)$ and in space of 
corresponding topological charges a vector structure.
For this one needs:

1) a vector basis $\{{\bf e}_i\}$;

2) a metris or/and a scalar product in this space.

In this basis all topological charges must have integer-valued components,
and a scalar product will determine interaction of different charges.
$$
{\bf q} = {\bf q}_1 + {\bf q}_2, \quad
q^2 = q_1^2 + q_2^2 + 2({\bf q}_1{\bf q}_2).
$$
In general, this metrics is not fixed, but is defined by properties 
of the initial theory.

%\newpage

For $T^n$ a basis $\{{\bf e}_i\}$ is the canonical one with
$$
({\bf e}_i{\bf e}_j)= \delta_{ik} = g_{ik}.      
$$

%\begin{picture}(300,50)
%\multiput(80,0)(0,10){6}
%{\line(1,0){90}}
%\multiput(80,0)(10,0){10}
%{\line(0,1){50}}
%\multiput(200,0)(10,0){10}
%{\line(1,5){10}}
%\multiput(200,0)(2,10){6}
%{\line(1,0){90}}
%\put(120,20){\vector(0,1){10}}
%\put(120,20){\vector(1,0){10}}
%\put(234,20){\vector(1,0){10}}
%\put(234,20){\vector(1,5){2}}
%\put(60,20){$T^n$}
%\put(260,20){$T_L$}
%\end{picture} 

%\bs
As a result the topological charges, corresponding to different $S^1_i,$ 
do not interact! Thus, the theories with ${\cal M} = T^n$ reduce from
topological PT point of view to the case ${\cal M} = S^1$. 
Moreover, one can show that
all tori $T_L = {\mbb R}^n/{\mbb L}$,
where ${\mbb L}$ is some full $n$-dimensional lattice in ${\mbb R}^n$ 
$$ 
{\mbb L} = \sum_{i=1}^n n_i {\bf e}_i, \,\, n_i \in {\mbb Z}_i, \,\, 
{\bf e}_i \in \{{\bf e}_i\}_L, \quad g_{ik}=\sum_{a=1}^n e^a_i e^a_k,
$$
($\{{\bf e}_i\}_L, \,i=1,...,n$ forms a basis of lattice ${\mbb L}$) 
reduce in this sence to the case $S^1$ also. It means that for all
systems, having such tori as a vacuum space, the TPT will have the 
same properties.

At the same time topological excitations on ${\cal M}$ with a rank of
$\pi_1({\cal M})$ $n\ge 2$ can have new additional properties. In particular,
they can form knotted configurations in case $n=2.$ For example, a "trefoil"
knot corresponds to ${\bf e}= (2,3)$. 

%\newpage

\bs

{\bf 2. Degenerated (or constrained) tori.}

\bs

For obtaining nontrivial generalizations one must to put on 
some constraints on tori $T^n$ and/or on the initial theory.
There is a regular method for obtaining such constraints (Bulgadaev 1996)
which may be usefull in different
regions of physics due to its relation only with theory of 
simple Lie group $G.$

%\newpage
%\bs

\underline {\bf Proposal:}

\bs

\fbox{\parbox[c][1.2\height]{12cm}{
One must consider theories connected with  the Cartan tori $T_G,$
the maximal abelian subgroups of the simple compact Lie groups $G.$
}}

\bs

Tori $T_G$ generalize a torus $T^n=T_{U(n)}$.  They all can be considered 
as appropriately constrained 
torus $T^N$ with large enough $N$. The corresponding GL type theories
can be written out in abstract general form in terms of group $G$,
but in the real physical systems their forms and corresponding 
symmetry breaking must be determined in each case
separately, since the underlying group structure is usually hidden.

\underline{Two examples}. 

Let us represent a torus $T^N$ in a form of diagonal matrix
$$
T^n = diag(\exp(i\varphi_1), ..., \exp(i\varphi_n)) =
$$
$$
diag(\exp(i({\bf e}_1 \mb {\bm $\varphi$}), ..., 
\exp(i({\bf e}_n \mb {\bm $\varphi$})),
$$
where a vector $\mb {\bm $\varphi$} = (\varphi_1,...,\varphi_n)$ is introduced.

Then
 
1) a condition $\sum_1^{n+1} \varphi_i = 0$ for torus $T^{n+1}$ 
gives a torus
of group $G=A_n=SU(n+1)$ in fundamental vector representation with
vectors $\{{\bf e}_a\},\; a =1,...,n+1,$ directed from origin to the 
vertices of hypertetrahedron in $n$-dimensional space; 

2) a reduction of torus $T^{2n}$ to the  form 
$\varphi_i = - \varphi_{n+i}, i=1,...,n,$  giving an additional 
reflection symmetry
$\varphi_i \to  - \varphi_i, i=1,...,n,$ reproduces a torus of group $D_n =
SO(2n)$ in fundamental vector representation with vectors 
$\{{\bf e}_a\}, \; a= 1,...,2n,$ 
directed to the vertices of $n$-dimensional hyperoctahedron.
 
In general, the Cartan torus $T_G,$ the maximal abelian
subgroup of group $G,$ consists of elements
$$
{\bf g} = e^{2\pi i({\bf H} \mb {\bm $\phi$})},\;
{\bf H} = \{H_1,...,H_n\} \in {\cal C}, \; [H_i,H_j] = 0 ,
$$
where $n$ is a rank of $G,$
${\cal C}$ is a maximal commutative Cartan
subalgebra of the Lie algebra ${\cal G}$ of the group $G.$ 
We assume here that
$({\bf H} \mb {\bm $\phi$})$ is usual euclidean scalar product.
Due to their commutativity, all $H_i$ can be diagonalized simultaneously.
Their eigenvalues,  called the weights  ${\bf w},$ depend on the concrete
representation of $G$ and ${\cal C}$ .
The weights $\{{\bf w}_a \}_{\tau}, \,
a = 1,..., p$, belonging to a $p$-dimensional irreducible
representation $\tau(G),$ form a set "of quantum numbers"  of this
representation.

All possible weights  ${\bf w}$ of the simply connected
group $G$ (or of the universal covering group $\ti G$ of
non-simply connected group $G$)
form a lattice of weights $L_{w}$. Its basis can be
chosen in different ways.
 In some cases the more convenient basis is a basis
of the vectors with minimal norms.

The weight lattice $L_{w}$, its reciprocal dual root lattice $L_v,$ and 
related  root lattice
$L_r$    are needed for finding
out $\pi_1(T_G).$  The last can
depend on the concrete representation $\tau (G).$ In two above examples
the vectors directed to the vertices of regular polyhedra form the sets of
weights of the corresponding fundamental representations, while their edges 
form the sets of roots (in this case the roots and dual roots coincide).

Since in any irreducible representation
$\tau(G)$ of dimension $p,$ one can choose the eigenvectors $|a>$ of
${\bf H}$ as a basis
$$
{\bf H}|a> =  {\bf w}_a |a>,\; a=1,...,p,
$$
then in this basis all $H_i$ (and any element ${\bf g}\in T_G$) have diagonal
form
$$
{\bf g}_{\tau}=
diag(e^{2\pi i({\bf w}_1 \mb {\scrs \bm $\phi$})},...,
e^{2\pi i({\bf w}_p \mb {\scrs \bm $\phi$})})
$$
Here some weights can coincide, or even some weights can be equal to 0.

The main differences of
this form from the usual representation of  $T_L$ type tori
are:

1) a dimension of diagonal matrices 
coincides with dimension $p$ of $\tau$-representation,
which is usually larger, than rank of $G$;

2) the set of weights
$\{{\bf w}\}_{\tau}$ has a discrete Weyl (or crystallographic) symmetry, 
which ensures
Weyl invariance of $T_G$ and results in the next two
properties
$$
\sum_{a=1}^p {\bf w}_a = 0, \quad
g_{ik} = \sum _{a=1}^p w^a_i w^a_k = B_{\tau}\delta_{ik},
$$
where constant $B_{\tau}$ depends on representation.
In case of $T_G$ the effective metrics $g_{ik}$  is
proportional to the euclidean one.

It is obvious, that in this representation all ${\bf g}\in T_G$
are periodic with a lattice of periods ${\mbb L}_{\tau}^{t}$,
inverse to the lattice $L_{\tau}$, generated
by weights ${\bf w}_a (a=1,...,p)$ of $\tau$-representation. It means that
$L_{\tau}^{t}$ form a set of all topological charges of
$\tau$-representation of $T_G.$ 

The  lattice $L^t_{\tau}$ satisfies the next restriction
$$
L_{w^*} \supseteq L^t_{\tau} \supseteq L_v.
$$
where $L_{w^*}$ is a weight lattice  of dual group $G^*.$
For $\tau = min$ a lattice
$L^t_{\tau} = L_v,$  for $\tau =ad$ a lattice $L^t_{\tau} = L_{w^*}.$
The lattices $L_v$ and $L_{w^*}$ differ by a factor, which is isomorphic
to the centre $Z_G$ of group $G$
$$
L_{w^*} / L_v = Z_G.
$$  
Thus the set $\{{\bf q}\}$ can vary from
the set of minimal vectors (it forms the so called Voronoi region or
Wigner - Seitz cell of the corresponding lattice) of the weight lattice
till that of the root lattice.  All possible cases are determined by
subgroups of the centre $Z_G.$ For groups $G$ with $Z_G=1$ the lattices
$L_v$ and $L_{w^*}$ coincide.

\bs

\fbox{\parbox[c][1.2\height]{12cm}{
When $L^t_{\tau} = L_w$ (or $L^t_{\tau} = L_{w^*}$) one can reproduce
all weights (i.e. quantum numbers) of group $G$ (or $G^*$) as the vector
topological charges of vortices!}}

\bs

In this relation becomes obvious the important role of tori $T_G,$
connected with self-dual lattices (they are $T_{U(n)}$  and $T_{E_8}$): 
all quantum numbers of corresponding 
groups have topological interpretation in terms of topological charges of
vortices. 
When $L^t_{\tau} \in L_w,$ one can enlarge $L^t_{\tau}$ by factorization of
$T_G$ over finite group $Z_{\tau} = L_w/L_{\tau}.$

%\newpage
\bs

\cl{\large {\bf III. $\sigma$-models on $T_G$, duality and effective
theories}}

\bs

The
euclidean
two-dimensional nonlinear $\sigma$-models on $T_G$,
generalizing
 $\sigma$-model on a circle $S^1,$ has the following
form
$$
{\cal S} = \frac{1}{2\alpha} \int d^2x Tr_{\tau}({\bf t}^{-1}_{\nu}
{\bf t}_{\nu}) =
\frac{(2\pi)^2}{2\alpha} \int d^2x Tr_{\tau}({\bf H}
{\mb {\bm $\varphi$}}_{\nu})^2
$$
$$
= \frac{(2\pi)^2}{2\alpha}B_{\tau}\int d^2x ({\mb {\bm $\varphi$}}_{\nu})^2,
$$
where ${\mb {\bm $\varphi$}}_{\nu} = \partial_{\nu} \mb {\bm $\varphi$},
\, \nu = 1,2.$ 
It will be convenient below to include a factor $B_{\tau}$ 
as a normalization
factor into trace $Tr_{\tau}.$ This gives us a canonical euclidean metric
in space of topological charges.

These theories are invariant
under direct product of right (R) and left (L)
groups $N_G^{R(L)},$ which are a semi-direct product of $T_G$ and $W_G$
$$
N_G = T_G \times W_G.
$$
The group $N_G$ is called a normalizator of $T_G$ and is a symmetry group
of torus $T_G.$

NSM on $T_G$ are the multicomponent generalization of $XY$-model, having
properties analogouos to those of $XY$-model:

1) a zero beta-functions $\beta(\alpha)$ due to flatness of $T_G;$

2) non-trivial homotopy group $\pi_1$ and corresponding vortex
solutions.

The general $N$-vortex solutions have the next form 
$$
\mbox{\bm $\varphi$}({\bf x}) = \sum_{i=1}^N {\bf q}_i\fr{1}{\pi}
\arctan(\fr{y-y_i}{x-x_i}),
$$
$$
{\bf q}_i \in L^t_{\tau}, \quad ({\bf q}_i {\bf w}_a)\in {\mbb Z},
\quad (x,y)\in {\mbb R}^2.
$$
Here $L^t_{\tau}$ is a lattice of all possible topological charges 
in $\tau$-representation.
The energy of $N$-vortex solution, $E_N,$ with a whole topological
charge $\sum _{i=1}^N {\bf q}_i = 0$ is
$$
E_N= \sum_i E^0_{q_i} + E_{Nint}, \quad
E^0_{q_i}= \fr{1}{2\alpha}C(a)({\bf q}_i{\bf q}_i), \quad
$$
$$
E_{Nint}=
\fr{2\pi}{2\alpha}\sum_{i\ne k}^N ({\bf q}_i{\bf q}_k)
\ln \fr{|x_i-x_k|}{a},
$$
where $E^0_{q_i}$ is a  "self-energy" (or an energy of the core) of
vortex with topological charge ${\bf q}_i$
 Only such solutions give
finite contribution to partition function of the theory.

Analogous topological excitations with  vectorial charges exist
also in the models of two-dimensional crystall melting (Nelson 1978). 

In quasi-classical
approximation (or in low T expansion) one can represent a partition
function of the $\sigma$-model on $T_G$
$$
{\cal Z}= \int D \mb {\bm $\varphi$} \exp(-{\cal S}[\mb {\bm $\varphi$}])
$$
as a grand
partition function of classical neutral Coulomb gas of vortex 
solutions with
minimal isovectorial topological charges
${\bf q}_i \in \{{\bf q}\}_{\tau},$ where
$\{{\bf q}\}_{\tau}$ is a set of minimal vectors of lattice of all possible
topological charges in $\tau$-representation $L^t_{\tau}$
$$
{\cal Z}= {\cal Z}_0 {\cal Z}_{CG},\quad
{\cal Z}_{CG}= \sum_{N=0}^{\infty} \frac{\mu^{2N}}{N!} \sump_{\{{\bf q}\}}
{\cal Z}_N(\{{\bf q}\}|\beta).
$$
Here 
${\cal Z}_0$ is a partition function of free massless isovectorial 
boson field
which corresponds to "spin waves" of $XY$-model
$$
{\cal Z}_0 = \int D \mb {\bm $\phi$}\exp (-{\cal S}_0[\mb {\bm $\phi$}]),
$$
$\sump$ in ${\cal Z}_{CG}$ goes over all neutral  configurations 
of minimal charges
${\bf q}_i \in \{{\bf q}\}_{\tau}$
with $\sum_1^N {\bf q}_i = 0,$
$$
{\cal Z}_N(\{{\bf q}\}|\beta)=
\prod_{i=1}^{N} \int d^2x_i \exp(-\beta H_N(\{{\bf q}\}))
$$
$$
H_N(\{{\bf q}\})= \sum_{i<j}^{N} ({\bf q}_i{\bf q}_j)D(x_i-x_j),
$$
$$
D(x)= \int \frac{d^2k}{(2\pi)^2}(e^{i({\bf k}{\bf x})}-1)f(ka)/k^2
\mathrel{\mathop {\sim}\limits_{|x|\gg a}} \frac{1}{2\pi} \ln|x/a|
$$
where
$$
\mu^2 = a^{-2} y^2 (det)^{-1/2}, \quad y^2 = e^{-E^0_q}
$$
is a chemical activity of Coulomb gas,
$det$ is a determinant of the fluctuations
over one vortex solution (futher we will suppose that it is equal to some
constant of order O(1) and assume that $det =1$),
$$
\beta = 4(\pi)^2/\alpha,
$$
$f(ka)$ is a regularisator such that
$$
lim_{k \to 0} f(ka)=1, \quad lim_{k \to \infty} f(ka) =0.
$$

There is a connection between compact $\sigma$-models on $T_G$ and
noncompact generalized SG theories
$$
{\cal Z}_{CG} = \int D\mb {\bm $\phi$} e^{-{\cal S}_{eff}}, \quad
{\cal S}_{eff} = \int \fr{1}{2\beta}(\partial \mb {\bm $\phi$})^2
-\mu^2 V(\mb {\bm $\phi$}),
$$
$$
V(\mb {\bm $\phi$}) =
\sum_{\{{\bf q}\}} \exp i({\bf q} \mb {\bm $\phi$}).
$$
where $\sum_{\{{\bf q}\}}$ goes over the set of minimal topological
charges, and $\mb {\bm $\phi$} \in {\mbb R}^n$.
Strictly speaking, the effective theories with arbitrary parameters $\mu$
and $\beta$ are more general than initial $\sigma$-models.

The account of vortices reduces
the initial symmetry group $N_G$ into discrete dual group
$W_{G^*} \times L_q^{-1}$ 
($W_{G^*}$ is a dual Weyl group, $L_q^{-1}$ is
a periodicity lattice   of potential $V$).
% and, consequently,
%is reciprocal to all ${\bf q} \in \{{\bf q}\}.$ It follows from their
%definitions that $L_q^{-1} =  L_{\tau}.$ 
This dual group generalizes
the dual group $Z_2 \times {\mbb Z}$  of $XY$-model.

Thus, in this semiclassical and long wavelength approximation

\bs

\fbox{
\parbox[c][1.2\height]{12cm}{{\it Compact}
theory on a torus $T_G$ with {\it continuous} symmetry $N_G$ appears
equivalent (again modulo ${\cal Z}_0$)
to {\it noncompact}
theory with periodic potential and an {\it infinite discrete} symmetry.}} 

\bs

%The potentials $V$ contain the sum over all minimal vectors $\{{\bf q}\}$
%and can coincide with characters of some representations of group $G.$
%For example, in case
%of $L^{-1}_{\tau} = L_v$ the sum in $V$ goes over all dual minimal
%roots.  
 
%\bs

For simply laced groups the potentials $V$ for groups from series $A,D,E$
($A_n = SU(n+1), D_n = SO(2n), E = E_{6,7,8}$) coincide
with characters of adjoint representations of these groups (modulo some
constant, corresponding to zero weight). In this case the general
theories can describe systems  with symmetry $G$ broken to
$N_G$.

The noncompact theories  can be considered also as corresponding
linear $\sigma$-models. For this reason

\bs 

\fbox{
\parbox[c][1.2\height]{12cm}{
{\it Compact nonlinear} $\sigma$-models on $T_G$ are
equivalent in this approximation to {\it noncompact linear} $\sigma$-models
on Cartan tori of dual group $T_{G^*}.$}} 

\newpage

%\bs

\cl{\large {\bf IV.  Phase diagram, critical properties and symmetries}}

\bs

The BKT type PT can be investigated  by renormalization of 
the corresponding effective field theories 
(Wiegmann 1978, Bulgadaev 1981,1996a). 

Under renormalization transformations both parameters $\mu$ and
$\beta$ are renormalized. It is convenient to introduce two
dimensionless parameters
$$
(\mu a)^2=g, \, \delta = \frac{\beta q^2 - 8\pi}{8\pi}
$$
where $q^2$ is a square of the norm of the minimal vector topological
charges from $\{{\bf q}\}.$ The effective theories are renormalizable
only if the vectors from  $\{{\bf q}\}$ belong to some lattice (here
$L^t_{\tau}$). A new critical properties can appear only if a geometry
of $\{{\bf q}\}$ is such that each vector ${\bf q} \in \{{\bf q}\}$
can be represented as a sum of two other vectors from $\{{\bf q}\}$.
It means that a potential $V$ has the next property
$$
V^2 = \sum_a \sum_b \exp (i({\bf q}_a +{\bf q}_b, \mb {\bm $\phi$})) =
$$
$$
\theta_0 + \theta V + \sum_{a+b \ne c} 
\exp (i({\bf q}_a +{\bf q}_b,\mb {\bm $\phi$})),
$$
where $\theta_0$ is doubled number of vectors in $\{{\bf q}\},$ having
their opposite vectors, 
$\theta$ is a number of times how each vector ${\bf q}$ can be 
represented as
a sum of two other vectors from $\{{\bf q}\},$ 
the last term contains vectors with norm larger than $|q|$ 
(so called higher harmonics). 
An appearence of $V$ on the right hand is very restrictive 
(the angles between vectors can be equal only to $\pi/3, 2\pi/3$
or $\pi/2$) and 
concides with a definition
of the root systems $\{{\bf r}\}$ of simple groups from series
$A,D,E$   or
with a definition of the root set of the even integer-valued
(in some scale) lattices of $\mbb {A,D,E}$ types. The sets of
minimal roots (or minimal dual roots) of all simple groups belong to
four series of integer-valued lattices $\mbb {A,D,E,Z}.$ For theories with
sets $\{{\bf q}\} \notin \mbb {A,D,E,}$ all critical properties will be
the same as for theories with $\{{\bf q}\} \in {\mbb Z}^n$.

The RG equations for $\{{\bf q}\}$ from these (i.e. $G=A,D,E$) lattices
have the next form 
$$
\fr{dg}{dl}=-2\delta g + B_{G}g^2, \quad
\fr{d\delta}{dl}= - C_{G}g^2.
$$
Here $B_{G}=\pi\theta_{G},$
and $C_{G}=2\pi K_G$, where $K_G$ is the value of the second Casimir
operator in adjoint representation (where ${\bf w}_a = {\bf r}_a$, the roots)
$$
\sum_{a} r^a_i r^a_j= K_G \delta_{ij}.
$$
The RG equations of such type with coefficients corresponding to the case
$G= A_2$ have been obtained firstly by Nelson (1978) under investigation of the
melting of the two-dimensional triangle lattice.

The eigenvalue of the second Casimir
operator $K_G$  for groups from
series $A,D,E$ can be expressed in terms of the corresponding Coxeter
number $h_G$
$$
K_G=2h_G, \quad h_G=
\frac{\mbox{(number of all roots)}}{\mbox{(rank of group)}},
$$
$$
h_{A_n} = n+1,\; h_{D_n} = 2(n-1),\; h_{E_{6,7,8}}= 12,18,30.
$$ 
This definition of the Coxeter number coincides with that of the Coxeter
number of the corresponding integer-valued lattices from series 
$\mbb {A,D,E}.$
The coefficient $B_G$ 
can be expressed also through the Coxeter number
$$
\theta_G = K_G - 4 = 2(h_G-2).
$$

\bs
 
\fbox{
\parbox[c][1.2\height]{12cm}{
Thus, all coefficients of RG equations depend only on
the Coxeter number $h_G$ or on the second Casimir value $K_G$.}}

\bs

The RG equations have two separatrices
with next declinations ($u_1$ corresponds to the phase separation line).
$$
u_{1,2}= (1/\pi h_G, -1/2\pi).
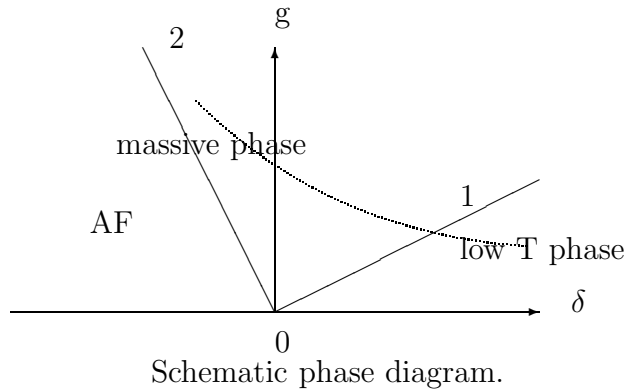
$$
\begin{picture}(400,150)(-75,0) 
\put(100,0){\vector(1,0){100}} 
\put(100,0){\vector(0,1){100}} 
\put(100,0){\line(-1,0){100}} 
\put(100,0){\line(2,1){100}}
\put(100,0){\line(-1,2){50}} 
\put(170,20){low T phase} 
\put(40,60){massive phase}
\put(30,30){AF} 
\put(210,0){$\delta$} 
\put(100,110){g} 
\put(100,-15){0}
\put(60,100){2}
\put(170,40){1}
\qbezier[100](70,80)(120,30)(195,25)
\end{picture} 
 
%\vspace{1cm} 
\bs
\cl{Schematic phase diagram.}

\vspace{0.5cm} 
 
 The dashed line of the initial values  corresponds to the initial
$\sigma$-model. This line is defined by the dependencies of the parameters
$\beta$ and $\mu$ on coupling constant $\alpha$. 

The critical exponent $\nu_G,$ determining divergence of the correlation
length $\xi$
$$
\xi \sim a\exp(A\tau^{-\nu_G}), \quad \tau = \fr{T-T_c}{T_c},
$$
is the inverse of the Lyapunov index of the separatrix 1 and equals
$$ 
\nu_G = 2/(2+h_G).  
$$ 
Since $h_G$ is a relatively rough characteristic, $\nu_G$  and other critical 
properties of different systems can coincide. 
 
All possible values of exponent $\nu_G$ are the next (Bulgadaev 1983,1996a)
 
\bs 
 
%\centerline{Table 1} 
 
$$ 
\ba{|c|c|c|c|c|c|c|c|c|c|} 
\hline 
{} & {} & {} & {} & {} & {} & {} & {} & {} & {}\cr 
G & A_n & B_n & C_n & D_n & G_2 & F_4 & E_6 & E_7 & E_8\cr 
{} & {} & {} & {} & {} & {} & {} & {} & {} & {}\cr 
\hline 
{} & {} & {} & {} & {} & {} & {} & {} & {} & {} \cr 
\nu_G & \fr{2}{n+3} & \fr{1}{n} & \fr{1}{2} & \fr{1}{n} & \fr{2}{5} 
& \fr{1}{4} & \fr{1}{7} & \fr{1}{10} & \fr{1}{16} \cr 
{} & {} & {} & {} & {} & {} & {} & {} & {} & {} \cr 
\hline 
\ea 
$$ 

\bs 
%\vspace{0.5cm} 
 
%The exponents $\nu_{A_1} = \nu_{D_2} = \nu_{B_n}$ correspond to the 
%initial KT exponent $\nu = 1/2.$  For $V( \phi)$ containing the set 
%of minimal roots $\{{\bf r}_a\}$ the exponents $\nu_G$  for groups 
%$B_n$ and $C_n$ pass into one another, since their root sets are 
%mutually dual. For other groups exponents remain the same. 
%Since $\nu_G$ depends only on $h_G$, they can coincide 
%for different groups having different rank and acting in different spaces. 
%This fact could be important when potential $V(\vec \phi)$ is composed 
%f the characters of different representations of different groups. 
 
The series $A_n$ gives a maximal set of possible values of $\nu_G:$ 
 $1/k$ and  $2/(2k+1)$ (where $k$ are integers 
$\ge 2$). 

\bs 
%\newpage

\underline {Low-T  phase properties}

\bs

In the low-temperature phase the correlation functions  of the fields 
exponentials equal to the free correlation functions with a renomalized 
"temperature" $\bar \beta$ which depends on initial values $\beta_0,$ 
$\bar \beta = lim_{l\to \infty} \beta(l)$: 
$$ 
\left< \prod_{s=1}^{t} \exp(i({\bf r}_s \mb {\bm$\phi$}(x_s)))\right> = 
\prod_{i\ne j}^{t}
 \left|\fr{x_i-x_j}{a}\right|^{\bar \beta ({\bf r}_i{\bf r}_j)/2\pi}, 
$$
$$ 
\sum_{i=1}^{t} {\bf r} = 0.  
$$ 
At the PT point (where  $\bar \beta = \beta^{*} = 8\pi/r^2 = 4\pi$) an 
additional logariphmic factor, related with the "null charge" behaviour 
of $g$ and $\delta$ on the 
critical separatrix  
(the phase separation line), 
appears in them: 
$$ 
\prod_{i\ne j}^{t} \left(\ln \left|\fr{x_i-x_j}{a}\right| 
\right)^{\beta^{*}({\bf r}_i{\bf r}_j)/2\pi A_G}= 
$$
$$
\prod_{i\ne j}^{t} \left(\ln \left|\fr{x_i-x_j}{a}\right| 
\right)^{h_G \cos({\bf r}_i{\bf r}_j)}, 
$$ 
where $A_G= 4/h_G$ is a coefficient in RG equations for $\delta$ on 
the critical separatrix 1. 

In this phase additional vortices can move like particles in dielectric media.

Free-like behaviour of the low-temperature phase 
(except logariphmic corrections at criticality) 
gives a possibility to use for their description conformal 
field theories with \underline {integer} central charges $C = n$, instead of 
PT points of two-dimensional systems with discrete symmetries, which 
are described by conformal theories with \underline {rational} 
central charges 
(Belavin et al. 1984, Dotsenko and Fateev 1984,1985). 
The BKT  PT can be considered as some 
degeneration of II order PT: it corresponds to the limiting case $k\to \infty$
(where $k$ is a level) of the
sequence of the restricted SG models, describing minimal conformal theories
(Smirnov and Reshetikhin 1989, LeClair 1989). Analogously, 
TPT in $\sigma$-models on $T_G$ are the limiting cases of unitary minimal 
conformal theories, connected with conformal $W$-algebras 
(Fateev and Lukyanov 1988).
In this relation it is interesting to note that there exists a puzzling
coincidence of  $\nu_G$ with "screening" factor in formulas for 
central charges 
of the affine Lie algebras $\hat {\mathfrak{G}}$ (Kac 1990) at level $k=2$
(though $T_G$ corresponds to $k=1$) 
$$ 
C_k = \frac{k}{k+h_G} dim G 
$$ 
and of coset realization of the
minimal unitary conformal models (Goddard et al. 1985,1986) at level $k=1$ 
$$ 
C_k = r\left(1-\frac{h_G(h_G +1)}{(k+h_G)(k+h_G +1)}\right). 
$$ 
 
\newpage 
  
\underline {Properties of massive phase}

\bs

In this phase all additional vector charges will be shielded like in plasma.
 
 A separatrix 2 
is a boundary of the asymptotically free (AF) region in UV-limit.
There is also another possibility of the enhancement of the symmetry of
the initial nonlinear  $\sigma$-model
on this separatrix.  $\sigma$-model has at classical level
two symmetries:
1) scale (or conformal) symmetry, 2) isotopic global symmetry
$N_G = T_G \times W_G.$
Conformal symetry is
spontaneously broken in IR region
by vortices. For this reason $\sigma$-model has in massive phase
a finite correlation length $\xi \sim  m^{-1}$, where  $m$ is a
characteristic mass scale of the theory. This mass must depend on 
the coupling
constant $\alpha$ or $\beta.$ 
 The behaviour of $m$ near PT point is
 described
by above formula, where
$$
\tau \sim \fr{\alpha - \alpha_c}{\alpha_c}.
$$
%There is another region in massive phase, the separatrix 2, where
%$m(\alpha)$ can be found.
%Since in IR-limit this separatrix attracts all trajectories in massive
%(or high-temperature) phase, it is very important to know an effective
%mass scale on it.

On separatrix 2 one obtaines
$$
m \sim\ \Lambda \exp{(-1/2\pi gh_G)}, \quad \Lambda \sim  a^{-1}.
$$
%The numerical factor in $\beta$-function can vary
%in dependence on a normalization of the coupling constan
%but 
The fact that on a separatrix 2 a mass scale is defined by $K_G \sim h_G$ is
a corollary of the independence of its
declination  on $G.$

This expression 
 coincides with those for chiral models on groups
(Polyakov 1975, Ogievetskii, Wiegmann, 1986) and with those, obtained
from an exact solution of the appropriate fermion theories in main
approximation on $g$ (Destri, De Vega 1987)
$$
m \sim \Lambda \exp (-2\pi/(gK_G/2)).
$$
Thus,
\bs

\fbox{\parbox[c][1.2\height]{12cm}{
  Mass scale on separatrix 2 coincides
(at least for $G=A,D,E$)
with that for all $G$-invariant theories
(chiral and fermionic), connected with simple Lie groups $G.$
}}

\bs

It means a possible restoration of the full symmetry group $G$
in $\sigma$-models with reduced symmetries $N_G$.
This can
be seen also from the equivalence of the effective field theories 
in  cases $G=A_{n-1} = SU(n), G = D_n = O(2n),$ to the
fermion theories with the same glodal symmetry groups $G$ (Coleman 1975,
Mandelstam 1975, Bulgadaev 1981, 1983)

%From here it follows, that in massive phase of chiral theory
%on $T_G$ ($G=A,D,E$) in minimal representation (when $L^t_{min} = L_v$)
There is strong dependence
of the index $\nu(\alpha)$,
which interpolates between $\nu = \nu_G$ near the PT point and $\nu =1$
near the separatrix 2.
The first region corresponds to symmetry
of $T_G,$ which is a torus normalisator $N_G= T_G \times W_G,$ while
the second one corresponds to more symmetric,  $G$-invariant,
situation.

Another interesting application of TPT in NSM on $T_G$ is connected with
string theory and its cosmological aspects. In this case TPT can be considered
as effective decompactification PT (Bulgadaev 1998), a massive phase 
corresponds to a compactified phase and a massless phase corresponds 
to a decompactified
phase.

\bs

%\newpage

\cl{\large {\bf V. d-dimensional conformal $\sigma$-model}}
\bs
\cl{\large {\bf and topological excitations}}

\bs

1. \underline {d-dimensional generalization}.

\bs 

Importance of topological exitations with logariphmically 
divergent energies in 1d systems has been earlier discovered 
by Anderson, Yuval and Hamann  in 1970, in papers
devoted to the Kondo problem. It was shown there that PT 
takes place in 1d Ising system with long-range  
interaction 
$$
J(r) \sim 1/r^2
$$ 
due to the presence of logariphmically 
interacting domain walls.   
 
The  next  properties unify these two 
models and give
the conditions of existence of 
topological exitations with logariphmic 
energy in d-dimensional NS models:

1) conformal invariance at classical level,

2) homotopical group 
$\pi_{d-1}({\cal M})$ must be nontrivial abelian discrete.

The first property defines form of action $S$ 
and the second one defines partially a topology of ${\cal M}$ 
$$ 
S=\frac{1}{2\alpha} 
\int d^d x d^d x' \psi_a (x)\boxtimes_{ab}^{(d)}(x-x')\psi_b (x'),
$$ 
where $\psi \in {\cal M}, \; a,b = 1,2...,n, \; n$ is a dimension of 
${\cal M}$ and a
form of kernel $\boxtimes$ depends on dimension of space $d$. 
For decoupled  internal and physical spaces  
$\boxtimes$ can be decomposed 
$$ 
\boxtimes_{ab}^{(d)}(x)= g_{ab} \Box_d (x),  
$$ 
where $g_{ab}$ is the Euclidean  metric 
of the space $\mathbb{R}^{N(n)}$, 
in which a manifold ${\cal M}$ can be embedded.  In the momentum space, 
for small $k$ 
$$ 
\Box_d (k)\simeq |k|^d (1 + a_1 (ka) + ...), 
$$ 
where $a$ is a UV cut-off parameter. 
Action ${\cal S}$ can be named $d$-dimensional {\emph{conformal}} 
nonlinear $\sigma$-model. 
The kernel $\Box_d$ generalizes an usual  local and 
conformal kernel of two-dimensional $\sigma$-model 
$$ 
\Box(k)\equiv \Box_2(k) = k^2 . 
$$ 
For local models an expression for $\boxtimes$ can be defined in 
terms of manifold ${\cal M}$ only 
$$ 
\boxtimes(x) = g_{ab}(\phi)\Box \delta(x)  
$$ 
In odd dimensions $\Box_d$ is nonlocal 
$$ 
\Box_d(x)\sim 1/|x|^{2d}, 
$$ 
but the models with such kernels are used often in physics. 

The simplest spaces with such properties are the spheres $S^{d-1}$. 
In 3d case this is a conformal (or van der Waals, 
since $\Box_3(x) \sim 1/|x|^6$) 
NS-models on sphere $S^2.$ 

An existence of topological excitations with logarithmic energy 
follows from the invariance of the kernel $B$ and  
the simplest topologically nontrivial 
excitations ${\bf n}_i = x_i/r$ under scale and conformal transformations 
$$
x_i \to  x'_i = x_i/r^2, \quad r \to r' = 1/r, \quad x_i/r = x'_i/r',
$$
$$ 
d^d x \to d^d x/|{\bf x}|^d,\quad
\fr{1}{|{\bf x}_1-{\bf x}_2|^{2d}} \to 
\fr{|{\bf x}_1|^d \;|{\bf x}_2|^d}{|{\bf x}_1-{\bf x}_2|^{2d}}, 
$$
and, consequently,
$$
{\cal S} \sim \int d^d x_1 d^d x_2 \; \fr{({\bf x}_1{\bf x}_2)}{r_1\;r_2}
\fr{1}{|{\bf x}_1-{\bf x}_2|^{2d}} 
$$
is invariant and dimensionless.
By direct calculation one can show that
$$
{\cal S} [{\bf n}] \approx C_d \ln \fr{R}{a}
$$
where $C_d = \fr{16}{\alpha}$, and different excitations interact 
through potential $G(r),$  inverse to $\Box_3(x),$ which has a logariphmic 
behaviour at large distancies
$$
G(r) \sim \ln \fr{r}{R}.
$$

\bs

2. \underline {Multicomponent generalization for $d>2$}.

\bs

A simple generalization of the sphere $S^{d-1}$, analogous to the torus $T^n$ 
in 2d case, is a bouquet of $n$ spheres
$$
B_n^{d-1} = S^{d-1}_1\vee... \vee S^{d-1}_n
$$
and all spaces ${\cal M}$ with this first nontrivial topological cell 
complex.
But the corresponding vector topological charges will not interact as in 
a case of torus. For obtaining an interaction of topological charges in 
3d case one needs to consider
NS-models on deformed $B^n$. In particular, NS-models 
on the maximal flag spaces $F_G=G/T_G$ of the simple compact 
groups $G$, with $\pi_2(F_G)= \mathbb {L}_v,$
(note that a sphere $S^2$ is a particular case of $F_G$: 
$S^2 = SU(2)/U(1)$)
will also have topological excitations with interacting vector 
topological charges ${\bf Q} \in \mathbb{L}_v$ and  logariphmic energy. 

Thus, these excitations will have the mixed properties of the two-dimensional
vortices and instantons: their topology is connected with $\pi_2$,
but they interact as vortices.

\bs

\cl{\large {\bf VI. Conclusions}}
 
\bs
  
1. It is shown that one must consider deformed tori for obtaining 
the interacting vector topological charges.

2. Vector topological charges form a lattice and in some cases can 
reproduce all quantum numbers of the corresponding groups.

3. A sequence of approximately equivalent 
transformations of 2d  models is constructed
\bs

\fbox{\parbox[c][2\height]{12cm}{\hspace{1cm} General GL Theory $\to$
NS Model $\to$
TopExcGas 

\cl{$\to$ General SG Theory}}}

\bs
It simplifies a problem and extracts all necessary long-wave properies 
of these theories! Here last theory has a pure group-theoretical structure
and is universal for  whole class of theories with the same symmetries.
 
4. All critical properties are classified by integer-valued lattices
from series $\mbb {A,D,E,Z}$ and are characterised by the corresponding
Coxeter numbers.

5. The possible scenarios of the dynamical enlargening of the initial 
internal
symmetry groups is proposed.

6. D-dimensional conformal NS-models, having the topological excitations 
with logarithmic energies, are constructed.

\bs

I would like to thank the organizers of the workshop for the invitation and 
the opportunity to give this talk. 

%\newpage


\bbib{100}
\bibitem{}  Anderson P.W., Yuval G.,Hamann D.R.,
Phys.Rev. {\bf B1} (1970) 4464.
\bibitem{}   Belavin A.A., Polyakov A.M., Schwartz A.S.,
Tyupkin Yu.S. {\bf 59B} (1975) 85.
\bibitem{}   Belavin A.A.,Polyakov A.M., Zamolodchikov A.B.,
Nucl.Phys. {\bf B241} (1984) 333.
\bibitem{}  Berezinskii V.L., JETP {\bf 59} (1970) 907,
{\bf 61} (1971) 1144.
\bibitem{}  Bulgadaev S.A., Phys.Lett. {\bf 86A} (1981) 213; 
Theor.Math.Phys. {\bf 49} (1981) 7.
\bibitem{}  Bulgadaev S.A., Nucl.Phys. {\bf B224} (1983) 349.
\bibitem{}  Bulgadaev S.A., JETP Letters {\bf 39} (1984) 264.
\bibitem{}  Bulgadaev S.A., JETP Letters {\bf 63} (1996)a 780;
hep-th/9906091.
\bibitem{}  Bulgadaev S.A., JETP Letters {\bf 63} (1996) 796;
hep-th/9901035.
\bibitem{}  Bulgadaev S.A., hep-th/9811226, (1998).
\bibitem{}  Coleman S., Phys.Rev., {\bf D11} (1975) 2088.
\bibitem{}  Destri C., de Vega H.J., Preprint CERN-TH, 4895/87.
\bibitem{}  Dotsenko Vl.S.,Fateev V.A.,
Nucl.Phys.{\bf B240} (1984) 312, {\bf B251} (1985) 691.
\bibitem{}  Fateev V.A.,Lukyanov S.L.,
Int.Jour.Mod.Phys. {\bf A13} (1988) 507.
\bibitem{}  Goddard P.,Kent A.,Olive D.,
Phys.Lett.{\bf B152} (1985) 88,
Commun.Math.Phys. {\bf 103} (1986) 105.
\bibitem{}   Green M., Schwarz J.H., Witten E., Superstrings Theory,
Cambridge, 1988, vol.1,2.
\bibitem{}  t'Hooft G., Nucl.Phys.{\bf B79} (1974) 276.
\bibitem{}  Jose J.,Kadanoff L., Kirkpatrick S.,Nelson D.,
Phys.Rev. {\bf B16} (1977) 1217.
\bibitem{} Kac V.N., Infinite dimensional Lie algebras, Cambridge
University Press, 1990.
\bibitem{}  Kosterlitz J.M.,Thouless J.P., J.Phys.{\bf C6} (1973) 118.
\bibitem{}  Kosterlitz J.M., J.Phys.{\bf C7} (1974) 1046.
\bibitem{}  LeClair A., Preprint PUPT-1124 (1989).
\bibitem{}  Legget A.J. et al. Rev.Mod.Phys. {\bf 59} (1987) 1.
\bibitem{}  Mandelstam S., Phys.Rev. {\bf D11} (1975) 3026.
\bibitem{} Nelson D.R., Phys.Rev. {\bf B18} (1978) 2318.
\bibitem{} Ogievetskii E., Wiegmann P.B., Phys.Lett. {\bf 168B} (1986) 360.
\bibitem{}  Polyakov A.M., Pisma v ZETP {\bf 20} (1974) 430.
\bibitem{}  Polyakov A.M., Gauge Fields and Strings, Harwood Academic
Publishers, 1987.
\bibitem{}  Popov V.N., unpublished (1971); Feynman integrals in quantum field theory
and statistical mechanics, Moscow. Atomizdat. 1976.
\bibitem{}  Reshetikhin N., Smirnov F., Commun.Math.Phys.,
{\bf 31} (1990) 157.
\bibitem{} Rice T.M., Phys.Rev.{\bf 140} (1966) 1889.
\bibitem{}  Schmid A., Phys.Rev.Lett. {\bf 51} (1983) 1506.
\bibitem{}  Schon G.,Zaikin A.D., Phys.Rep. {\bf 198} (1990) 237.
\bibitem{}  Stanley H.E., Kaplan T.A., Phys.Rev.Lett., {\bf 17} (1966)
913.
\bibitem{} Toulouse G., Kleman M., J.Physique Lett. {\bf 37} (1976) 149.
\bibitem{} Volovik G.E., Mineev V.P., JETP {\bf 45} (1977) 1186
\bibitem{} Wiegmann P.B., J.Phys.{\bf C11} (1978) 1583.
%\bibitem{}  Wiegmann P.B., Tsvelik A.,
%Advances in Phys. {\bf 32} (1983) 453.
\ebib
\end{document}